\documentclass{eptcs}
\usepackage{url}
\usepackage{breakurl}
\usepackage{graphicx}
\title{Development of a Translator from LLVM to ACL2\footnote{Approved
    for Public Release, Distribution Unlimited}}
\author{
David S. Hardin\footnote{The views expressed are those of the authors 
and do not reflect the official policy or position of the Defense
Advanced Research Projects Agency (DARPA) or the U.S. Government.}\\
\institute{Advanced Technology Center\\Rockwell Collins\\
Cedar Rapids, IA, USA}
\email{david.hardin@rockwellcollins.com}
\and
Jennifer A. Davis\\
\institute{Advanced Technology Center\\Rockwell Collins\\
Cedar Rapids, IA, USA}
\email{jennifer.davis@rockwellcollins.com}
\and
David A. Greve\\
\institute{Advanced Technology Center\\Rockwell Collins\\
Cedar Rapids, IA, USA}
\email{david.greve@rockwellcollins.com}
\and
Jedidiah R. McClurg\footnote{The work described herein was performed
  during a co-op session at Rockwell Collins.}\\
\institute{Department of Computer Science\\University of Colorado\\
Boulder, CO, USA}
\email{jedidiah.mcclurg@colorado.edu}
}

\def\titlerunning{Development of a Translator from LLVM to ACL2}
\def\authorrunning{D. Hardin, J. Davis, D. Greve, \& J. McClurg}

\begin{document}
\maketitle

\begin{abstract}

In our current work a library of formally verified
software components is to be created, and assembled, using the 
Low-Level Virtual Machine (LLVM) intermediate form, into
subsystems whose top-level assurance relies on the assurance of the
individual components.  We have thus undertaken a project to build a
translator from LLVM to the applicative subset of Common Lisp accepted
by the ACL2 theorem prover.  Our translator produces executable ACL2
formal models, allowing us to both prove theorems about the translated
models as well as validate those models by testing.  The resulting
models can be translated and certified without user intervention, even
for code with loops, thanks to the use of the \texttt{def::ung} macro
which allows us to defer the question of termination.  Initial
measurements of concrete execution for translated LLVM functions
indicate that performance is nearly 2.4 million LLVM instructions per
second on a typical laptop computer.  In this paper we overview the
translation process and illustrate the translator's capabilities by
way of a concrete example, including both a functional correctness
theorem as well as a validation test for that example.

\end{abstract}


\section{Introduction}\label{intro}

In our current work, we need to create formally verified software
systems from a library of verified components assembled using the
Low-Level Virtual Machine (LLVM) intermediate form \cite{LLVM}.  To
accomplish this we have undertaken a project to build a translator
from LLVM to the applicative subset of Common Lisp
\cite{CommonLispHyperSpec} accepted by the ACL2 theorem prover
\cite{ACL2book}, and perform verification of the software system using
ACL2's automated reasoning capabilities.

LLVM is the intermediate form for many common compilers, including the
\texttt{clang} compiler used by Mac OS X and iOS developers.  LLVM
supports a number of language frontends, and LLVM code generation
targets exist for a wide variety of machines, including both CPUs and
GPUs.  LLVM is a register-based intermediate language in Static Single
Assignment (SSA) form \cite{SSA}.  As such, LLVM supports any number
of registers, each of which is only assigned once, statically
(dynamically, of course, a given register can be assigned any number
of times).  Andrew Appel has observed that ``SSA form is a kind of functional
programming'' \cite{SSAfun}; this observation, in turn, inspired us to
build a translator from LLVM to the applicative subset of Common Lisp
accepted by the ACL2 theorem prover.  Our translator produces an
executable ACL2 specification that is able to support proof-based 
verification, as well as validation via testing.

\section{Toolchain Overview}

\begin{figure}
\begin{center}
\includegraphics[scale=0.7]{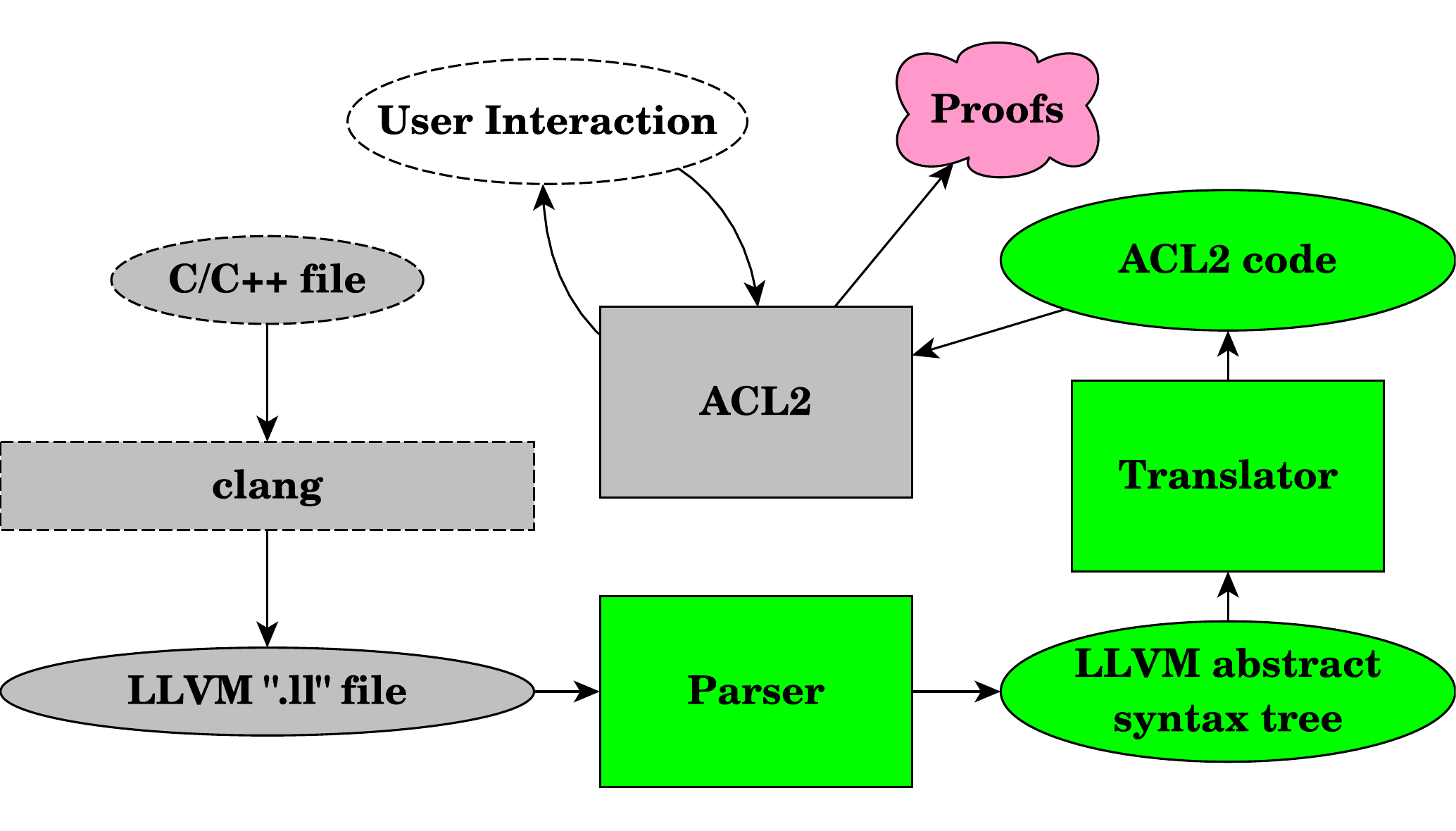}
\end{center}
\caption{LLVM-to-ACL2 translation toolchain.}
\label{toolchain}
\end{figure}

Our translation toolchain architecture is shown in Figure
\ref{toolchain}.  The left side of the figure depicts a typical
compiler frontend producing LLVM intermediate code.  LLVM output can
be produced either as a binary ``bitcode'' (.bc) file, or as text (.ll
file).  We chose to parse the text form, producing an abstract syntax
tree (AST) representation of the LLVM program.  Our translator
converts the AST to an ACL2 model of the code which ACL2 is able to
certify automatically.  Once certified, the model can be loaded
together with conjectures that one wishes to prove about the code.  In
addition to proving theorems about the translated LLVM code, ACL2 can
also be used to execute test vectors against the translated model at
reasonable speeds.

The translator is written in OCaml \cite{OCaml}.  OCaml was chosen
because the translator developer was proficient in that language.  The
translator comprises less than 9500 lines of OCaml code, and  
employs a parser generator whose input file, describing the LLVM 
grammar, is some 1800 lines.  The translator 
successfully parses all 5000+ legal \texttt{.ll} files in
the LLVM source distribution.  The translator produces an AST from
the input, removes aliases, extracts functions from labelled basic
blocks, constructs parameter lists, determines declaration order, 
then generates the ACL2 code for each function.

\section{An Example}\label{example}

As an example, consider the C source code of Figure \ref{Ccode}.
This function counts the number of occurrences of a given value in the
first n elements of an array.  (NB: By default the \texttt{clang} compiler
treats all \texttt{int} values as 32 bits wide, and all \texttt{long}
values as 64 bits wide.)

\begin{figure*}
\begin{verbatim}
unsigned long occurrences(unsigned long val, unsigned int n, 
                          unsigned long *array) {
  unsigned long num_occur = 0;
  unsigned int j = 0;
  for (j = 0; j < n; j++) {
    if (array[j] == val) num_occur++;
  }
  return num_occur;
}
\end{verbatim}
\hrulefill
\caption{Example C code to count occurrences of an input value in an array.}
\label{Ccode}
\end{figure*}

This is one of the more simple examples in the regression test suite
for our framework, and it fails to exercise many of the more advanced 
features of our translator and proof infrastructure.  However, its relative simplicity
allows us to narrate a complete translation and verification within 
the confines of this paper.

We produce the LLVM code for this function by invoking \texttt{clang}
as follows: \texttt{clang -O4 -S -emit-llvm occurrences.c}.  The
generated LLVM code for clang version 4.2 (which supports LLVM 3.2) is
excerpted in Figure \ref{LLVMcode}.

\begin{figure*}
\begin{verbatim}
define i64 @occurrences(i64 %val, i32 %n, i64* %array) {
  %1 = icmp eq i32 %n, 0
  br i1 %1, label %._crit_edge, label %.lr.ph

.lr.ph:
  %indvars.iv = phi i64 [ %indvars.iv.next, %.lr.ph ], [ 0, %0 ]
  %num_occur.01 = phi i64 [ %.num_occur.0, %.lr.ph ], [ 0, %0 ]
  %2 = getelementptr inbounds i64* %array, i64 %indvars.iv
  %3 = load i64* %2, align 8, !tbaa !1
  %4 = icmp eq i64 %3, %val
  %5 = zext i1 %4 to i64
  %.num_occur.0 = add i64 %5, %num_occur.01
  %indvars.iv.next = add nuw nsw i64 %indvars.iv, 1
  %lftr.wideiv = trunc i64 %indvars.iv.next to i32
  %exitcond = icmp eq i32 %lftr.wideiv, %n
  br i1 %exitcond, label %._crit_edge, label %.lr.ph

._crit_edge:
  %num_occur.0.lcssa = phi i64 [ 0, %0 ], [ %.num_occur.0, %.lr.ph ]
  ret i64 %num_occur.0.lcssa
}
\end{verbatim}
\hrulefill
\caption{LLVM code for the occurrences example.}
\label{LLVMcode}
\end{figure*}

Observe that LLVM output is similar to assembly code, with labels and
low-level opcodes like \texttt{br} (branch), \texttt{icmp} (integer
compare) and \texttt{load} (load from memory).  Registers are
prepended with the ``\%'' character, and are given
sometimes-meaningful names.  Consistent with the SSA philosophy, no
register appears on the left hand side of an assignment (``='') more
than once.  A peculiar feature of LLVM code is the \texttt{phi}
instruction, which provides register renaming at a branch target.  We
will use the \texttt{phi} in our ACL2 translation to match formal to
actual parameters, as will be detailed later.

\section{Translation to ACL2}\label{translation}

Treating the SSA as a functional program, we convert register assignments
into nested let bindings.  Each label is treated as a unique function, so we
produce \texttt{defun} forms for \texttt{@occurrences},
\texttt{.lr.ph}, and \texttt{.\_crit\_edge}.  The formal parameters
for the functions can be determined by consulting the left hand side
of the \texttt{phi} functions; thus, \texttt{.\_crit\_edge} should
have a formal parameter \texttt{\%num\_occur\_dot\_0\_dot\_lcssa} in
its parameter list (accounting for the differences in allowed
characters in parameter names).  We also need to identify parameters
that are read, but not modified --- most of our functions will thus
require \texttt{\%val}, \texttt{\%n} and \texttt{\%array} as input
parameters.

We are left, then, with the question of how to translate memory and
memory transactions.  Typically in ACL2, a machine state data
structure is declared, and passed as a parameter to all functions that
read and/or write elements of the state.  If a given function updates
the state, the modified state must be returned.  Obviously, for a large
state, functional update of the state can become quite expensive.  
In an earlier version of the translator \cite{LLVMtoACL2}, we utilized an ACL2
single-threaded object (stobj) \cite{STOBJ} to represent state.  The 
destructive update property of stobjs provided good performance 
when executing translated functions on concrete state.  
However, stobjs are not currently compatible with the deferred 
termination technology that we wished to employ, so stobjs were 
abandoned in favor of typed records \cite{defrecord03}.

\subsection{LLVM State Representation}\label{state}

The raw LLVM state is represented using the following
\texttt{defstructure} (\texttt{defstructure} can be found in
\texttt{data-structures/structures.lisp} in the ACL2 community books):

\begin{verbatim}
(defstructure raw-st
  (retval (:assert (natp retval) :rewrite ))
  (stack  (:assert (usb32 stack) :rewrite ))
  (frame  (:assert (usb32 frame) :rewrite ))
  (mem    (:assert (wf-mem  mem) :rewrite ))
  (:options :guards
            (:predicate stp)
            (:keyword-updater update-raw-st)))
\end{verbatim}

This declaration indicates that the LLVM state is composed of four
fields: \texttt{retval}, \texttt{frame}, \texttt{stack}, and
\texttt{mem}.

The LLVM memory model assumes a 32-bit underlying architecture.  In
other words, the primitive \texttt{int} type is 32 bits wide, as are
memory addresses.  The \texttt{mem} field contains a memory 
of 8-bit words, implemented as an association list.  While the size of
the memory array is, technically,
unbounded, as a practical matter, its size is limited to 4 GB since we
use 32-bit addresses.  The memory could easily be upgraded to 64-bit
addressing in the future.  Little-endian byte ordering is assumed by
the low-level \texttt{rd.n} and \texttt{wr.n} primitives, but
big-endian could also be readily supported.
 
The \texttt{retval} field is used to return values from procedure
calls.  This was needed historically since all translated functions
accept state and return only state.  In the future this field could 
be replaced by a stack lookup or a multiple-valued return.  Note that the size of
retval is unbounded (a \texttt{natp}).  This allows us to return
objects such as structures by converting them to natural numbers.

The \texttt{stack} field is a 32-bit pointer that identifies the next
available stack location in memory.  The stack grows towards infinity.
The \texttt{frame} field is a 32-bit pointer that identifies the top
of the previous stack frame.

\subsection{Completing the Translation}

The translated function for the LLVM code beginning at label
\texttt{.\_crit\_edge} is depicted below.  The
translator ensures that the state record, \texttt{st}, is passed to,
and returned from, all translated functions, even simple ones such as
this.  Referring to the LLVM of Figure \ref{LLVMcode},
\texttt{.\_crit\_edge} does nothing more than stash the value of
\texttt{num\_occur} in the \texttt{retval} field of the state and
return the updated state.

\begin{verbatim}
(def::un occurrences_%_dot__crit_edge (%num_occur_dot_0_dot_lcssa st)
  (declare (xargs :signature ((i64_p stp) stp)))
  (let* ((st (update-retval %num_occur_dot_0_dot_lcssa st)))
    st))
\end{verbatim}

This definition employs features of Greve's \texttt{def} package,
provided as part of the ACL2 community books \cite{assumeterm} \cite{defung}.
The \texttt{def::un} macro, found in the \texttt{coi/util/defun} book, improves
upon ACL2 \texttt{defun} by providing both input and output ``type''
signatures.  In the example above, the signature

\begin{verbatim}
  (declare (xargs :signature ((i64_p stp) stp)))
\end{verbatim}

says that the given function takes two inputs, the first of which
satisfies the \texttt{i64\_p} predicate and the second of which
satisfies the \texttt{stp} predicate, and returns one output,
satisfying the \texttt{stp} predicate.  The \texttt{:signature} form
accepts all standard Common Lisp type declarations (i.e.:
\texttt{string}, \texttt{integer}, \texttt{unsigned-byte}) as well as
generic predicate symbols and even lambda expressions.  In the
\texttt{defun} form generated by \texttt{def::un} the input element
types become guards and the signature declaration is interpreted as a
theorem about the return type of the function, in this case:

\begin{verbatim}
(DEFTHM I64_P-STP-IMPLIES-STP-OCCURRENCES_%_DOT__CRIT_EDGE
  (IMPLIES (AND (I64_P %NUM_OCCUR_DOT_0_DOT_LCSSA)
                (STP ST))
           (STP (OCCURRENCES_%_DOT__CRIT_EDGE
                     %NUM_OCCUR_DOT_0_DOT_LCSSA ST)))
  :RULE-CLASSES
  (:REWRITE
    (:FORWARD-CHAINING
         :TRIGGER-TERMS ((OCCURRENCES_%_DOT__CRIT_EDGE
                              %NUM_OCCUR_DOT_0_DOT_LCSSA ST)))))
\end{verbatim}

This theorem states that, for a given invocation of the function, if
the inputs satisfy their associated predicates, then the value
returned by the function satisfies its predicate.  This rule is
written as a forward-chaining rule so that this ``type'' information
is readily propagated.  Such properties are
often needed and \texttt{def::un} saves us the drudgery of explicitly
writing rules of this form ourselves for each new function that we
define.

It should be noted that \texttt{def::un} is similar in spirit to the
\texttt{defunc} (``defun with contracts'') form in ACL2s \cite{ACL2s},
although the details of the respective contract specifications differ
(\texttt{defunc} features \texttt{:input-contract} and
\texttt{:output-contract} keywords, whose interpretation is somewhat
different from the \texttt{:signature} of \texttt{def::un}.  The 
\texttt{def::un} macro also allows a user to specify congruence relations
satisfied in each argument position, a capability which
\texttt{defunc} does not provide.

\subsection{Translating Loops}

Loops in LLVM are translated into recursive functions in ACL2.
Automatically generating recursive functions in ACL2 is challenging
because ACL2 requires the identification of a well-founded relation
for use as a measure to ensure that the recursion
terminates.  Without user input, however, 
automatically identifying such a measure for arbitrary LLVM code is 
impossible.  To alleviate this issue we employ the
\texttt{def::ung} macro.  This macro, found in
\texttt{coi/defung/defung.lisp} in the ACL2 community books, allows us to admit arbitrary
recursive functions without the need of a measure.  Rather, the macro
generates a companion domain predicate\footnote{The macro also
generates a measure function which is guaranteed to decrease with
each recursive call in the domain of the function.} which, when true,
ensures that the function terminates.  We thus exchange the need to
identify a measure before admitting the function for the inconvenience
of reasoning about the function's domain predicate during subsequent
proofs.  As an aside, the \texttt{def::ung} macro supports the same
set of \texttt{:signature} declarations as \texttt{def::un} and also
produces function definitions that can be executed efficiently.  The
\texttt{def::ung} macro, however, does not currently support
stobjs or multiple-value returns, limitations that have
driven several trade-offs in our translation framework.

The original C source in our example (see Figure \ref{Ccode}) used
a \texttt{for} loop, which can be naturally expressed as a while loop
(test at the top of the loop), but the
corresponding loop in LLVM (the \texttt{.lr.ph} block) is of do-while
form (test at the bottom of the loop).  Transforming loops into 
do-while form is standard for clang in
all but the least optimized mode of operation.  The high-level
specifications used in our work, however, employ while loops
exclusively.  To ease the process of relating our LLVM implementations
to their high-level specifications, our LLVM translator has been
designed to emit only while loops.

Another useful optimization is to isolate the potentially complex body
of a loop from the recursive call in the ACL2 output. Hence for any given loop in LLVM,
we emit a single ``step'' function containing the body of the
loop. This ``step'' function is called by the while function in ACL2
to advance the state.  During proof, the step function can 
oftentimes be disabled, thus simplifying and speeding the proof process.

Upon completion of a loop the next action performed is typically a
jump to a subsequent block of sequential code.  Having the base case
of a recursive function invoke an arbitrarily complex block of code,
however, complicates the reasoning process unnecessarily.  In our
translation, therefore, when the recursive function terminates, it
simply returns the live variables and the current state.  We
introduce a ``while\_wrap'' function that calls the while loop and
then calls the function corresponding to the next LLVM block to be
executed.

The general form of the generated ACL2 code for the ``Nth'' loop
of an arbitrary LLVM procedure named ``fun'' is shown in 
Figure \ref{outline}.  Note that the \texttt{mvlist} macro returns 
multiple values as a single list and the \texttt{metlist} macro
emulates multiple-value binding for functions returning such lists of
values.  These macros can be found in the book
\texttt{coi/util/mv-nth}.  We use these macros to
skirt the fact that \texttt{def::ung} does not support multiple-value
returns.

\begin{figure*}
\begin{verbatim}
(def::un fun_continue (... st)
  (declare (xargs :signature ((... stp) stp)))
  (let ((st <post-loop functionality>))
    st))

(def::un fun_step_N (... st)
  (let ((st <loop functionality>))
    (let ((done <set or clear done bit>))
      (mvlist done ... st))))

(def::ung fun_step_N_while (done ... st)
  (declare (xargs :signature ((natp ... stp) ... stp)))
  (if (= done 1) (mvlist ... st)
    (metlist ((done ... st) (fun_step_N ... st))
      (fun_step_N_while done ... st))))

(def::un fun_step_N_while_wrap (... st)
  (declare (xargs :signature ((... stp) stp)))
  (metlist ((... st) (fun_step_N_while 0 ... st))
    (let ((st (fun_continue ... st)))
      st)))

(def::un fun_N (... st)
  (declare (xargs :signature ((... stp) stp)))
  (let ((done <set or clear done bit>))
    (if (= done 1) (fun_continue ... st)
      (fun_step_N_while_wrap ... st))))
\end{verbatim}
\hrulefill
\caption{Outline of the generated ACL2 code for an LLVM loop.}
\label{outline}
\end{figure*}

Observe that, although one could never accuse this clique of 
generated functions of being terribly efficient (e.g., there are 
many more function calls and returns than one would like to see, 
with much marshalling/unmarshalling of data), nonetheless this 
structure preserves the tail recursive nature of the LLVM input, 
allowing us to perform computations on large 
input terms without exhausting the Lisp stack.  We will revisit 
performance issues in Section \ref{exec}.

The generated ``step'' function for the occurrences example 
is depicted in Figure \ref{step_0}.  Note that the last thing that
this step function does is return a list of parameters to be used
for the next execution of the step function.  Note particularly that
the first value in the returned list, corresponding to the position of
the \texttt{done} parameter, is set to the value of the
\texttt{\%exitcond} register, which indicates whether the updated
loop index variable is equal to the array size.  Also note that
throughout the remainder of the example, we
use the \texttt{bits} function from \texttt{rtl/rel9} in the ACL2 
community books whenever we need a result modulo some finite number of
bits.

\begin{figure*}
\begin{verbatim}
(def::un occurrences_step_0 (done %num_occur_dot_01 %indvars_dot_iv
                             %array %n %val st)
  (declare (xargs :signature ((natp i64_p i64_p _30_p i32_p i64_p stp)
                              natp i64_p i64_p _30_p i32_p i64_p stp)))
  (let* 
      ((%2 (+ %array (_30_gep (list %indvars_dot_iv))))
       (%3 (i64_frombytes (loadbytes *i64_size* %2 st)))
       (%4 (icmp= %3 %val))
       (%5 (zext %4 1 64))
       (%_dot_num_occur_dot_0
        (bits (+ %5 %num_occur_dot_01) 63 0))
       (%indvars_dot_iv_dot_next
        (bits (+ %indvars_dot_iv (bits 1 63 0)) 63 0))
       (%lftr_dot_wideiv (bits %indvars_dot_iv_dot_next 31 0))
       (%exitcond (icmp= %lftr_dot_wideiv %n)))
    (mvlist %exitcond %_dot_num_occur_dot_0 %indvars_dot_iv_dot_next
            %array %n %val st)))
\end{verbatim}
\hrulefill
\caption{Generated ACL2 code for the occurrences loop body.}
\label{step_0}
\end{figure*}

The top-level \texttt{occurrences} function, shown in Figure
\ref{top-level}, is defined in terms of \texttt{occurrences\_0},
the final function admitted by our loop translation.
\texttt{occurrences\_0} represents the start of the do-while
loop.

\begin{figure*}
\begin{verbatim}
(def::un occurrences (%val %n %array st)
  (declare (xargs  :signature (( i64_p i32_p _24_p stp) stp)))
  (let*
      ((st (init-stack-frame st))
       (st (begin-stack-frame st))
       (st (occurrences_0 %val %n %array st)))
    (end-stack-frame st)))
\end{verbatim}
\hrulefill
\caption{Generated top-level driver code for the translated occurrences example.}
\label{top-level}
\end{figure*}

\section{Concrete Execution}\label{exec}

It is advantageous to be able to validate the translated models by
running them against concrete inputs.  Since all of our functions are
executable, we can readily perform such validation testing.  In the ACL2 code
of Figure \ref{concrete}, we set up an initial state, in which the stack and frame
pointers are set to a high memory location so that the stack cannot
overwrite our data array (we place the array at address
\texttt{\#x8000}).  We then write a number of 64-bit values into
memory at increasing addresses, initializing the array.  This
state is then passed to the translated top-level \texttt{occurrences} function,
along with inputs for the \texttt{val} and \texttt{n} parameters, as
well as the location of the array in memory.  After the execution of
\texttt{occurrences}, the return value is fetched using the \texttt{retval}
accessor.

\begin{figure*}
\begin{verbatim}
(def::un occurrences-test1 ()
  (declare (xargs :signature (() natp)))
  (let*
      ((myst (st 0 #xffff0000 #xffff0000 nil))
       (myst (update-mem
             (wr.n 8 #x8038 399
               (wr.n 8 #x8030 234
                 (wr.n 8 #x8028 0
                   (wr.n 8 #x8020 75
                     (wr.n 8 #x8018 399
                       (wr.n 8 #x8010 399
                         (wr.n 8 #x8008 (1- (expt 2 64))
                           (wr.n 8 #x8000 20 (mem myst))))))))) myst)))
    (retval (occurrences 399 8 #x8000 myst))))
\end{verbatim}
\hrulefill
\caption{Concrete test case for the translated occurrences example.}
\label{concrete}
\end{figure*}

As we have written the value 399 into the array three times, when we
run \texttt{(occurrences-test1)} from the ACL2 prompt, it returns the
correct value: 3.

We need not, however, restrict ourselves to small arrays.  The
underlying typed record representation for memory does not explicitly
store ``0'' values; thus, any address that has not been explicitly
written with a non-zero value is assumed to have a value of 0.  Thus, 
we can replace the last line in the test of Figure \ref{concrete} by

\begin{verbatim}
    (retval (occurrences 0 1000000 #x8000 myst))
\end{verbatim}

This corresponds to one million executions of the occurrences LLVM inner
loop, plus some start-up and clean-up code (which we ignore here).
Execution of this modified test returned the correct result (999993),
and ACL2's \texttt{time\$} function returned 3.8 seconds of real time for the above on a late
2012 MacBook Pro.  Further, the occurrences LLVM inner loop consists of 9 LLVM
instructions; thus, \[(1,000,000 * 9) / 3.8 \approx 2,370,000 \mbox{ LLVM
instructions per second.}\]  This result is encouraging, considering that
the translated Lisp code has not yet been engineered for performance
in any serious way.

\section{Reasoning about Translated Functions}\label{reasoning}

In order to reason about a function such as \texttt{occurrences} in
ACL2, we first need to perform abstraction on the data types;
particularly, we wish to abstract the input array to a Lisp list.
This can be done with the aid of a ``lift'' function, as follows:

\begin{verbatim}
(def::ung liftlist (done j array n st)
  (declare (xargs :signature ((natp natp natp natp stp) nat-listp)))
  (if (equal done 1) nil
    (let* 
        ((ptr (+ array (* j 8)))
         (val (wfrombytes 8 (loadbytes 8 ptr st)))
         (j (bits (1+ j) 63 0))
         (done (if (equal (bits j 31 0) n) 1 0)))
      (cons val (liftlist done j array n st)))))
\end{verbatim}

Note that we are modeling our lift function in a manner that closely
reflects the behavior of our underlying LLVM implementation.  In
particular, we have chosen to admit our lift function using
\texttt{def::ung} rather than to admit it as a standard ACL2
definition with a measure.  While a reasonable measure for this
function does, in fact, exist, the measure for a similar function that,
for example, traversed a linked list rather than simply iterating through an array,
would likely not exist.  Thus, we proceed with this example as though
a measure for our function is not available.

The list-based specification of \texttt{occurrences} is fairly conventional:

\begin{verbatim}
(def::un occurlist (val list)
  (declare (xargs :signature ((natp nat-listp) natp)))
  (if (endp list)
      0
    (+ (if (= val (car list)) 1 0)
       (occurlist val (cdr list)))))
\end{verbatim}

We wish to prove that the translated \texttt{occurrences} function
operating over an array in memory produces a result equal to the
\texttt{occurlist} function operating over a (lifted) list.  In order to
carry out this proof, we must reason about the domain predicates
generated by \texttt{def::ung}, which can be quite tedious.  For
example, we need for the domain of \texttt{occurrences\_step\_0\_while}
to be the same as the domain of \texttt{liftlist}.  Unfortunately,
however, \texttt{occurrences\_step\_while} has an additional
argument.  In order for the domains to be the same, we need to show
that this argument is irrelevant.  We can show this by induction, but
first we need an appropriate induction scheme.  This scheme is
illustrated in \texttt{occurrences\_step\_0\_while-induction-2}.
Note that the admission of this scheme leverages the
\texttt{occurrences\_step\_0\_while-domain} and
\texttt{occurrences\_step\_0\_while-measure} functions introduced by
\texttt{def::ung}.

\begin{verbatim}

(defun occurrences_step_0_while-induction-2 (done num num2 %next
                                             %array %n %val st)
  (declare 
    (xargs :measure 
      (occurrences_step_0_while-measure done num %next
                                        %array %n %val st)))
  (if (not (occurrences_step_0_while-domain done num %next
                                            %array %n %val st))
      st
    (if (equal done 1)
        (mvlist  num %next %array %n %val st)
      (metlist ((done2 num2 %next2 %array2 %n2 st2)
                  (occurrences_step_0 done num2 %next
                                      %array %n %val st))
        (metlist ((done num %next %array %n %val st)
                    (occurrences_step_0 done num %next
                                        %array %n %val st))
          (occurrences_step_0_while-induction-2 done num num2 %next
                                                %array %n %val st))))))

\end{verbatim}

With the induction scheme in hand the proof of the irrelevance of the
second argument is trivial.

\begin{verbatim}

(defthm domains-equiv-num_occur-irrelevant-lemma
  (implies 
    (occurrences_step_0_while-domain done num_occur1
                                     j array n val st)
    (occurrences_step_0_while-domain done num_occur2
                                     j array n val st))
  :hints
    (("Goal" :induct
       (occurrences_step_0_while-induction-2 done num_occur1 num_occur2
                                             j array n val st)
           :in-theory (enable occurrences_step_0))))

\end{verbatim}

Knowing that the second argument is irrelevant allows us to show that
\texttt{occurrences\_step\_0\_while-domain} and
\texttt{liftlist-domain} are equivalent by mutual implication.
\begin{verbatim}

(defthm occurrences_step_0_while-implies-liftlist
  (implies (occurrences_step_0_while-domain done num_occur
                                            j array n val st)
           (liftlist-domain done j array n st))
  :hints (("Goal" :induct (occurrences_step_0_while done num_occur
                                            j array n val st)
           :in-theory (enable occurrences_step_0))))

(defthm liftlist-implies-occurrences_step_0_while
  (implies (not (occurrences_step_0_while-domain done num_occur
                                                 j array n val st))
           (not (liftlist-domain done j array n st)))
  :hints (("Goal" :induct (liftlist done j array n st)
           :in-theory (enable occurrences_step_0))))

\end{verbatim}

Once we know that the domains of the two functions are equivalent we
can prove that the LLVM loop satisfies its specification.    ACL2
is able to prove this theorem automatically because of the way that we 
constructed the \texttt{liflist} function.  (NB: the macro
\texttt{val} in this example extracts the number of occurrences
computed by the LLVM loop from the returned list of values.)

\begin{verbatim}

(defthm occurrences_rec_equiv--thm
  (implies
    (and (stp st) (natp done) (natp n) (natp array)
         (natp val) (bvecp val 64)
         (natp j) (natp num_occur) (bvecp num_occur 64))
     (equal (val 0 (occurrences_step_0_while done num_occur
                                             j array n val st))
            (bits (+ num_occur
                    (occurlist val (liftlist done j array n st)))
                  63 0)))
  :hints (("Goal" :induct
            (occurrences_step_0_while done num_occur
                                      j array n val st)
           :in-theory
            (enable occurrences_step_0_while occurrences_step_0))))

\end{verbatim}

The top-level specification is expressed in terms of \texttt{occurlist}:

\begin{verbatim}

(def::un occurrences_spec (val n array st)
  (declare (xargs :signature ((natp natp natp stp) natp)))
  (if (zp n)
      0
    (bits (occurlist val (liftlist 0 0 array n st)) 63 0)))

\end{verbatim}

The final equivalence theorem is then as follows:

\begin{verbatim}

(defthm occurrences_equiv--thm
  (implies (and (stp st) (natp n) (natp array) (bvecp val 64))
           (equal (retval (occurrences val n array st))
                  (occurrences_spec val n array st))))

\end{verbatim}

This result follows easily once we have proved \texttt{occurrences\_rec\_equiv--thm}.

\section{Related Work}

Zhao \emph{et al.} \cite{Vellvm} produced several different
formalizations of operational semantics for LLVM in Coq
\cite{CoqRefMan}, noting that their intention is to produce a verified
LLVM compiler, similar to the verified CompCert compiler due to Leroy
\cite{Leroy2009} (CompCert does not utilize the LLVM intermediate
form).  As such, their emphasis on formalizing LLVM operational
semantics makes sense.  We also considered creating an ``LLVM
interpreter'' in ACL2 (Zhao \emph{et al.} utilized the OCaml
extraction capability of the Coq environment to produce such an
interpreter), resulting in a ``deep embedding'', but decided that a
translation to ACL2 (thus producing a ``shallow embedding'') would allow
us to begin proving properties about LLVM programs with much less
effort.  Our approach was also influenced by Magnus Myreen's
``decompilation into logic'' work \cite{decomp}.  Our approach could
be characterized as a sort of decompilation into logic, but we do not
go to the same lengths as Myreen to assure that the decompilation
process is sound.  We also have the advantage of starting with a form
that is functional, whereas Myreen has tackled the much more difficult
problem of decompiling imperative machine code.

\section{Conclusion and Future Work}

We have built a translator from the LLVM intermediate form to the
applicative subset of Common Lisp accepted by the ACL2 theorem prover.
The translator produces executable tail-recursive ACL2 specifications,
and we have utilized this executability in order to validate our
translated models via testing.  Performance of concrete execution for
translated LLVM functions has been measured at nearly 2.4 million LLVM
instructions per second on a typical laptop computer.

Future work will focus on continued enhancements to the translator and
support books for enhanced automated reasoning, particularly dealing
with translations that produce mutual recursions.  The translator will
also need to be updated to support new LLVM releases.

\section{Acknowledgments}

We thank the anonymous referees for their helpful comments.
This work was sponsored in part by the United States Department of Defense.

\bibliographystyle{eptcs}
\bibliography{fm}
\end{document}